\documentclass[useAMS,usenatbib,fleqn]{mn2e}  
\usepackage{times}
\usepackage{graphicx}
\usepackage{amsmath}
\def\Msun{\hbox{$\rm\thinspace M_{\odot}$}}

\def\swift{{\it Swift}}
\def\xte{{\it RXTE}}
\def\gro{GRO~J1655$-$40}
\def\gx{GX~339$-$4}
\def\xteja{XTE~J1550$-$564}
\def\xtejb{XTE~J1650$-$500}

\def\fouru{4U~1543$-$475}
\def\h{H~1743$-$322}
\def\gsim{\mathrel{\hbox{\rlap{\hbox{\lower4pt\hbox{$\sim$}}}\hbox{$>$}}}}
\def\lsim{\mathrel{\hbox{\rlap{\hbox{\lower4pt\hbox{$\sim$}}}\hbox{$<$}}}}
   \title[Radiative efficiency in X-ray binaries]{Evidence for changes in the radiative efficiency of transient black hole X-ray binaries}

   \author[A. J. Eckersall et al.]{A. J. Eckersall$^{1}$\thanks{aje16@le.ac.uk},
        S. Vaughan$^{1}$, G. A. Wynn$^{1}$\\
        $^{1}$Department of Physics and Astronomy, University of Leicester, Leicester, LE1 7RH, UK}
\begin{document}

\date{Draft \today}
 
\pagerange{\pageref{firstpage}--\pageref{lastpage}} \pubyear{2015}

\maketitle

\label{firstpage}

\begin{abstract}
 We have used pointed \xte\ data to examine the long-term X-ray light curves of six transient black hole X-ray binaries during their decay from outburst to quiescence. In most cases, there is a period of exponential decay as the source approaches the soft- to hard- state transition, and another period of exponential decay following this transition as the source decays in the hard state. The e-folding times change around the time of the state transition, from typically $\approx 12$ d at the end of the soft state to $\approx 7$ d at the beginning of the hard state. This factor $\sim2$ change in the decay time-scale is expected if there is a change from radiatively efficient emission in the soft state to radiatively inefficient emission in the hard state, overlying an exponential decay in the mass accretion rate. This adds support to the idea that the X-ray emitting region is governed by radiatively inefficient accretion (such as an advection-dominated or jet-dominated accretion flow) during the fading hard state.
\end{abstract}

\begin{keywords}
   accretion, accretion discs -- black hole physics -- X-rays: binaries
\end{keywords}
%
%________________________________________________________________

\section{Introduction}
\label{sect:intro}

Low-mass X-ray binaries (LMXBs) comprise a black hole (BH) or neutron star (NS), often called the primary, that accretes gas from a secondary star (also called the companion, or donor) with a mass $M_2 \lsim 1.5 \Msun$. An accretion disc is formed around the BH when the secondary fills its Roche lobe and material passes towards the BH \citep{shakura73, mcclintock06, done07}. In this paper, we concentrate on LMXBs with a BH primary; these systems are powerful but transient sources of X-rays. They spend most of their time in quiescence, with very low luminosity in X-rays and other wavebands, interrupted by outbursts that usually last for weeks--months during which the X-ray luminosity can be comparable to the Eddington luminosity $L_{\rm Edd}$ ($\sim 10^{38} M_{\rm BH}/$\Msun\ erg s$^{-1}$).

The current explanation for the outburst cycle is based on the disc instability model \citep[DIM; e.g. ][]{vanparadijs96, king98, dubus01, lasota01}. In the quiescent state, the accretion disc gas is mostly neutral and has a low viscosity. The rate at which mass is transferred from the secondary to the disc is typically larger than mass accretion rate through the disc on to the BH, such that the disc mass and temperature increase over time. As the temperature approaches $\sim 10^4$ K hydrogen ionization causes a dramatic change in the opacity of the gas within the disc that is associated with an increase in the disc viscosity. This triggers an increase in the accretion rate on to the primary, causing an outburst. During the outburst, the disc (and related corona/jet) is a powerful source of X-rays. These X-rays irradiate the accretion disc and raise its temperature above that expected from viscous heating alone, keeping the disc in the hot, high-viscosity phase. Nevertheless, eventually the mass accretion rate will drop, reducing the level of X-ray irradiation and allowing the disc to return to its cool, low-viscosity state. At this point the source returns to quiescence.

One of the generic predictions of the DIM, after inclusion of X-irradiation of the outer disc, is a period of exponential decay in the mass accretion rate: $\dot{M} \propto \exp(-t / \tau_{\rm m})$ \citep{king98, dubus01, lasota01}. The duration and e-folding time $\tau_{\rm m}$ is controlled by system parameters (such as the disc size). Such exponential decays have been observed in several XRB outbursts \citep{chen97, kuulkers98, jonker10, stiele12, tomsick14}, although often the decays are interrupted by flares and other re-brightening events.

Within each outburst, XRBs usually show several distinct `states' characterized by particular X-ray spectral and timing behaviour, but also correlated with other behaviour such as radio emission \citep[e.g.][]{remillard06, done07, belloni10}. The two most commonly observed states are the \emph{soft} state and the \emph{hard} state. The soft state (sometimes known as the thermal-dominated state) is generally seen near the peak of the outburst and is characterized by an X-ray spectrum dominated by thermal emission from the accretion disc with a temperature $T_{\rm peak} \sim 1$ keV \citep{remillard06}, and a weak, non-thermal tail of emission to higher energies. The soft state shows only weak rapid X-ray variability and weak, if any, radio emission \citep{fender09}. The X-ray emission is powered by a luminous, radiatively efficient accretion disc that is assumed to extend in to the innermost stable circular orbit ($r_{\rm ISCO} = 6r_{\rm g}$ for a non-rotating BH, where the gravitational radius is $r_{\rm g} = GM/c^2$). By radiative efficiency we mean the fraction of the accreting mass converted to escaping radiation, $\eta = L_{\rm bol}/\dot{M}c^2$. The hard state -- always seen during the rise to outburst and decay to quiescence, and sometimes near outbursts peak -- is characterized by an X-ray spectrum dominated by a non-thermal (power law) component, extending up to energies $\sim 100$ keV, with a much weaker, cooler thermal component ($T_{\rm peak} \lsim 0.3$ keV) sometimes detectable at lower energies \citep{remillard06}. The radio and timing properties are also different: the hard state shows strong, rapid X-ray variability \citep{vanderklis06, belloni10} and persistent (flat spectrum) radio emission \citep{fender09}. The source of the non-thermal emission is not fully understood but is often thought to be the result of either inverse-Compton scattering (of soft photons emitted from the accretion disc) in a corona of hot electrons above the disc \citep{haardt93, dove97, poutanen98}, or synchrotron and self-Compton emission produced in the base of the radio jet \citep{markoff01, markoff05}. There are also intermediate states; these are usually short-lived and occur during transitions between the longer lived hard and soft states \citep{belloni10}. It remains an open question how these bright states are related to the quiescent state \citep{fender03, miller-jones11, plotkin13, fender14, reynolds14}

Two varieties of accretion flow models, with distinct X-ray emission mechanisms, are often used to explain the differences between the soft and hard states. One is the standard, optically thick and geometrically thin accretion disc with its strong thermal emission \citep{shakura73}. This disc (when extending down to $r_{\rm ISCO}$) produces the thermal emission that dominates the X-ray spectrum and is assumed to have a high radiative efficiency ($\eta \sim 0.1$, \citealt{shakura73}) that is approximately constant with accretion rate. We therefore expect a linear scaling between X-ray luminosity and mass accretion rate, $L_{\rm X} \propto \eta \dot{M} \propto \dot{M}$, at least in the regime where $\dot{M}$ is high enough to maintain a such a disc. The other type of accretion flow is any of a group of alternatives to the optically thick accretion disc. Examples of these are the advection dominated accretion flow \citep[ADAF;][]{narayan95} and the jet-dominated accretion flow \citep[JDAF;][]{falcke04}. These are optically thin and generate X-rays via Comptonization, bremsstrahlung and/or cyclo-synchrotron. In the non-thermal-dominated hard states the inner accretion disc may be replaced by this optically thin flow \citep[e.g.][]{esin97} which has a lower radiative efficiency that depends on the accretion rate, e.g. $\eta \propto \dot{M}$. This leads to a non-linear scaling of X-ray luminosity with accretion rate: $L_{\rm X} \propto \eta \dot{M} \propto \dot{M}^2$ \citep{narayan95}. See e.g. \citet{fender03}, \citet{merloni03}, \citet{kording06} and \citet{coriat11} for further discussion of the link between X-ray and radio luminosities, accretion rate and radiative efficiency. 

\citet{knevitt14} studied the difference in the period distribution of the known LMXBs with NS and BH primaries. They found a dearth of very short period BH systems relative to NS systems that could be reproduced if the BH systems are radiatively inefficient below some critical mass accretion rate. Shorter period binaries have a lower maximum luminosity and may be radiatively inefficient through a substantial fraction of their outbursts. \citet{knevitt14} compared different models for the switch between efficient and inefficient accretion and found that only a dramatic change, to an efficiency scaling $\eta \propto \dot{M}^{\beta}$ with $\beta \gsim 3$, could explain the observed lack of short-period BH systems.

When the mass accretion rate is decaying exponentially, and the emission is dominated by a radiatively efficient accretion disc, the X-ray luminosity should decay with the same e-folding time: $L_{\rm X} \propto \dot{M} \propto \exp(-t / \tau_{\rm m})$. If the emission is dominated by a radiatively inefficient flow, the X-ray luminosity will decay faster, e.g. for $\eta \propto \dot{M}$ we will have $L_{\rm X} \propto \dot{M}^2 \propto \exp(-2t / \tau_{\rm m})$. The e-folding time of the luminosity is shorter (e.g. $\tau_{\rm m}/2$). Therefore, as discussed in \citet{homanAtel2005}, if there is a change from a radiatively efficient to a radiatively inefficient accretion regime during the period of exponential decay in the mass accretion rate, there should be a decrease in the e-folding time of the X-ray luminosity during the decay. In this paper, we test this hypothesis using some of the best \xte\ monitoring data of XRB outbursts.

%______________________________________________________________

\section{Observations and Data Reduction}
\label{sect:obs}

\subsection{Sample definition}

The sample of XRB outbursts is shown in Table~\ref{tab:outburstdetails}. We selected all the confirmed and candidate transient BH XRBs listed in \citet{remillard06}. For each one we, identified all the long ($>$few days) outbursts based on daily light curves of each source from the \xte\ All-Sky Monitor, and where possible also used data from the \swift\ Burst Alert Telescope, and \emph{MAXI} (Monitor of All-Sky X-ray Image) Gas Slit Camera. These data alone were insufficient to accurately define the decay rates. We therefore selected the subset of outbursts with a sufficient number ($>20$) of pointed \xte\ observations to estimate decay rates in both soft and hard states. 

\subsection{\xte\ observations}

\begin{figure}
  \centering
  \includegraphics[width=6.5 cm, angle=90]{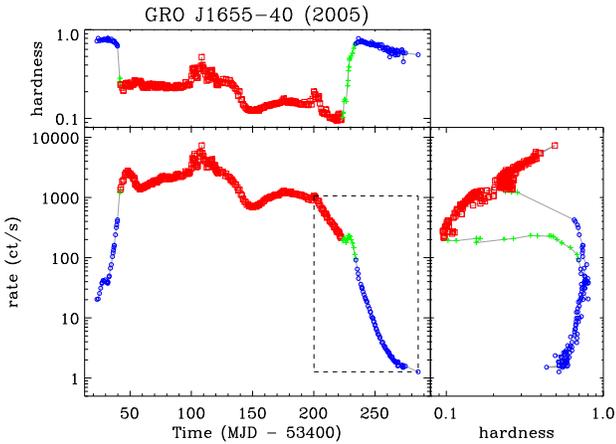}
  \caption{\xte\ light curve in the $\approx 3-20$ keV band showing the full 2005 outburst of \gro, along with a HID (right-hand panel) and hardness ratio curve (top panel). The plot has been coloured to show the spectral states of the source as the outburst progresses: blue circles for the hard state, green crosses for intermediate and red squares for soft. The dashed lines highlight the region shown in Fig. \ref{fig:gro_plot_zoom}.}
\label{fig:gro_plot}
\end{figure}

For each selected outburst we extracted pointed \xte\ data from the High Energy Astrophysics Science Archive Research Center (HEASARC) archive. We only took data from the Proportional Counter Array (PCA) instrument, in the {\tt STANDARD2}  mode which has $129$ energy channels. We extracted both source and background spectra for each observation, along with spectral response files using the standard \xte\ software {\sc heasoft} v6.15.1), and then computed background-subtracted count rates in three energy bands: $A$ = channels $4$-$44$ ($\approx 3$-$20$ keV), $B$ = channels $4$-$10$ ($\approx 3$-$6$ keV) and $C$ = channels $11$-$20$ ($\approx 6$-$10$ keV), as in \citet{motta12}. The background is obtained using models as \xte/PCA background cannot be measured. We used the hardness ratio $H = C/B$ \citep{homan05}. Using these we produced light curves, hardness ratio curves and hardness intensity diagrams \citep[HIDs;][]{belloni05} for each outburst. Fig.~\ref{fig:gro_plot} shows the data for the intensively observed 2005 outburst of \gro.

At low count rates ($<10$ ct s$^{-1}$), the data could be affected by systematic errors in the background subtraction and by contaminating sources in the PCA field of view, including diffuse Galactic emission. This often makes the decays appear to slow down towards quiescence \citep[e.g.][]{homanAtel2005,jonker10}. For each count rate, we computed errors based on standard Poisson statistics. However, these usually underestimate the random scatter in the light curves, as the dominant source of variance is systematic, not random, e.g. the intrinsic, rapid variability of the sources (particularly strong in the hard states) and systematic errors resulting from subtraction of the instrumental background (a more significant effect at low fluxes). As a way to assign similar statistical weight to data points that may differ in flux by two to three orders of magnitude, we assumed $3$\% errors for all count rates, in both soft and hard states. Thus, a fit is considered `good' if the model matches to within $\sim 3$ per cent for most data points.

\subsection{\swift\ Observations}

We have also made use of \swift\ X-ray Telescope (XRT) observations to track the outburst of \gro\ to lower count rates than is possible using only the \xte\ PCA data. As the XRT is an imaging telescope, it is less sensitive to background issues, so we use it to verify our \xte\ background subtraction. Of our sample only \gro\ has any useful observations. There are six \swift\ observations during the hard-state decay: three just after the end of the state transition back into the hard state at around MJD $53640$, and three a few weeks later at around MJD $53665$. The \swift\ XRT data were reduced using {\sc heasoft} v6.15.1. Photon event lists and exposure maps were produced using {\sc xrtpipeline} v0.13.1. For each observation, a source spectrum was extracted from photons in a circular extraction region of radius 30 arcsec, and a background spectrum was extracted from a source-free region of the same image. Response matrix files were generated using {\sc xrtmkarf} v0.6.1.

The energy bandpass and response of the \swift\ XRT differs and \xte\ PCA, and so we converted the \swift\ data to equivalent PCA rates for comparison with the \xte\ monitoring data. We used {\sc xspec} v12.8.2 \citep{arnaud96} to fit a model to the \swift\ XRT spectrum for each observation. The model comprised a blackbody plus power law, all modified by neutral absorption (the {\tt TBABS} model; \citealt{wilms00}). We extrapolated the best-fitting model for each observation to cover the PCA energy range and then used the PCA response matrix to convert this to an equivalent PCA count rate. The resulting PCA-equivalent count rates are shown on Fig.~\ref{fig:gro_plot_zoom2}.

%______________________________________________________________

\section{Analysis and Results}
\label{sect:analysis}

\subsection{The Model}
\label{sect:model}

If radiative efficiency is a power-law function of the central mass accretion rate, $\eta \propto \dot{M}^{\beta}$, and the mass accretion rate follows an exponential decay with e-folding time $\tau_{\rm m}$, the X-ray luminosity should follow $L_{\rm X} = A \exp(-t (1+\beta)/ \tau_{\rm m})$. We therefore fit a simple model $L_{\rm X} = A \exp(-t/\tau) + C$ to decaying intervals of the light curves to estimate the decay time-scales during soft-state and hard-state intervals to give luminosity decay timescales $\tau_{\rm s}$ and $\tau_{\rm h}$, respectively. The constant $C$ accounts for any constant light from contaminating background sources. 

Assuming the radiative efficiency is constant at high luminosities ($\beta = 0$ in the soft state), the luminosity decay in the soft state should track the decay in $\dot{M}$, i.e. $\tau_{\rm s} = \tau_{\rm m}$. But if the radiative efficiency scales with accretion rate ($\beta > 0$) in the hard state, the X-ray luminosity will have a different e-folding time, $\tau_{\rm h} = \tau_{\rm m} / (1+\beta)$. We can then estimate $\beta$  by comparing the decay times in soft and hard states as $\beta = \tau_{\rm s} / \tau_{\rm h} - 1$. If there is no change in radiative efficiency, we expect $\beta = 0$, while if there is a change to $\eta \propto \dot{M}$ regime we should see $\beta = 1$.

\subsection{Results}

We have studied the final weeks of all the outbursts listed in Table~\ref{tab:outburstdetails2}. Based on an initial, visual inspection of the light curves, for each source we identified two regions of decay separated by the soft--hard state transition. (During this transition -- spanning intermediate states -- the light curve often stays approximately constant or even rises slightly). The model above was fitted\footnote{The fitting was performed by minimizing the usual $\chi^2$ fit statistic.} to each of these regions to give estimates of $\tau_{\rm s}$ and $\tau_{\rm h}$. These are given in Table~\ref{tab:outburstdetails} and the data and models are plotted in the Appendix. All light curves are in the $3$-$20$ keV band.

\begin{table*}
  \centering
  \begin{tabular}{c c c c c c c c}
    \hline
    System name & Outburst & No. of obs. & $\tau_{\rm s}$ (d) & Decay length &$\tau_{\rm h} $ (d)& Decay length & $\beta$ \\
    (1) & (2) & (3) & (4) & (5) & (6) & (7) & (8) \\
    \hline
    \gro\ & $2005$ & $495$ & $14.02 \pm\ 0.13$& $50$ & $6.85 \pm\ 0.05$ & $73$ & $1.05 \pm 0.01$ \\
    \xteja\ & $1998$ & $234$ & $8.97 \pm\ 0.14$ & $21$ & $2.33 \pm\ 0.05$ & $9$ & $2.85 \pm\ 0.07$ \\
    & $1999$ & $66$ & $14.80 \pm\ 1.11$ & $8$ & $6.87 \pm\ 0.03$ & $70$ & $1.15 \pm\ 0.09$ \\
    \xtejb\ & $2001$ & $106$ & $16.18 \pm\ 0.25$ & $21$ & $-$ & $-$ & $-$ \\
    \gx\ & $2002$ & $174$ & $16.28 \pm\ 0.28$ & $13$ & $18.21 \pm\ 0.13$ & $37$ & $-0.11 \pm\ 0.01$ \\
    & $2004$ & $251$ & $17.09 \pm\ 1.29$ & $4$ & $14.15 \pm\ 0.07$ & $56$ & $0.21 \pm\ 0.02$ \\
    & $2007$ & $244$ & $16.94 \pm\ 0.34$ & $13$ & $7.43 \pm\ 0.32$ & $28$ & $1.28 \pm\ 0.06$ \\
    & $2010$ & $294$ & $-$ & $-$ & $11.46 \pm\ 0.28$ & $19$ & $-$ \\
    \fouru\ & $2002$ & $102$ & $6.10 \pm\ 0.15$ & $12$ & $4.97 \pm\ 0.04$ & $54$ & $0.23 \pm\ 0.01$ \\
    \h\ & $2003$ & $219$ & $14.06 \pm\ 0.37$ & $11$ & $4.81 \pm\ 0.07$ & $40$ & $1.92 \pm\ 0.06$ \\
    & $2004$ & $49$ & $22.06 \pm\ 0.82$ & $7$ & $5.66 \pm\ 0.14$ & $18$ & $2.90 \pm\ 0.13$ \\
    & $2005$ & $23$ & $12.92 \pm\ 1.28$ & $4$ & $5.49 \pm\ 0.12$ & $10$ & $1.35 \pm\ 0.14$ \\
    & $2009$ & $49$ & $20.07 \pm\ 1.68$ & $9$ & $8.02 \pm\ 0.17$ & $16$ & $1.50 \pm\ 0.13$  \\
    & $2010$ & $58$ & $16.92 \pm\ 0.86$ & $10$ & $6.95 \pm\ 0.11$ & $23$ & $1.43 \pm\ 0.08$ \\
    & $2011$ & $39$ & $11.75 \pm\ 0.52$ & $8$ & $7.39 \pm\ 0.15$ & $17$ & $0.59 \pm\ 0.03$  \\
    \hline
  \end{tabular}
  \caption{Details of the outbursts used in our sample. Column 1 gives the source name, column 2 gives the year of each outburst and column 3 gives the total number of \xte\ observations used. Columns 4 and 6 show the e-folding times, columns 5 and 7 show the number of observations included in the fit, and column 8 shows the $\beta$ values. Missing values indicate light curves with insufficient data to properly fit exponential curves.}
  \label{tab:outburstdetails}
\end{table*}

\begin{figure}
\centering
  \includegraphics[width=6.5 cm, angle=90]{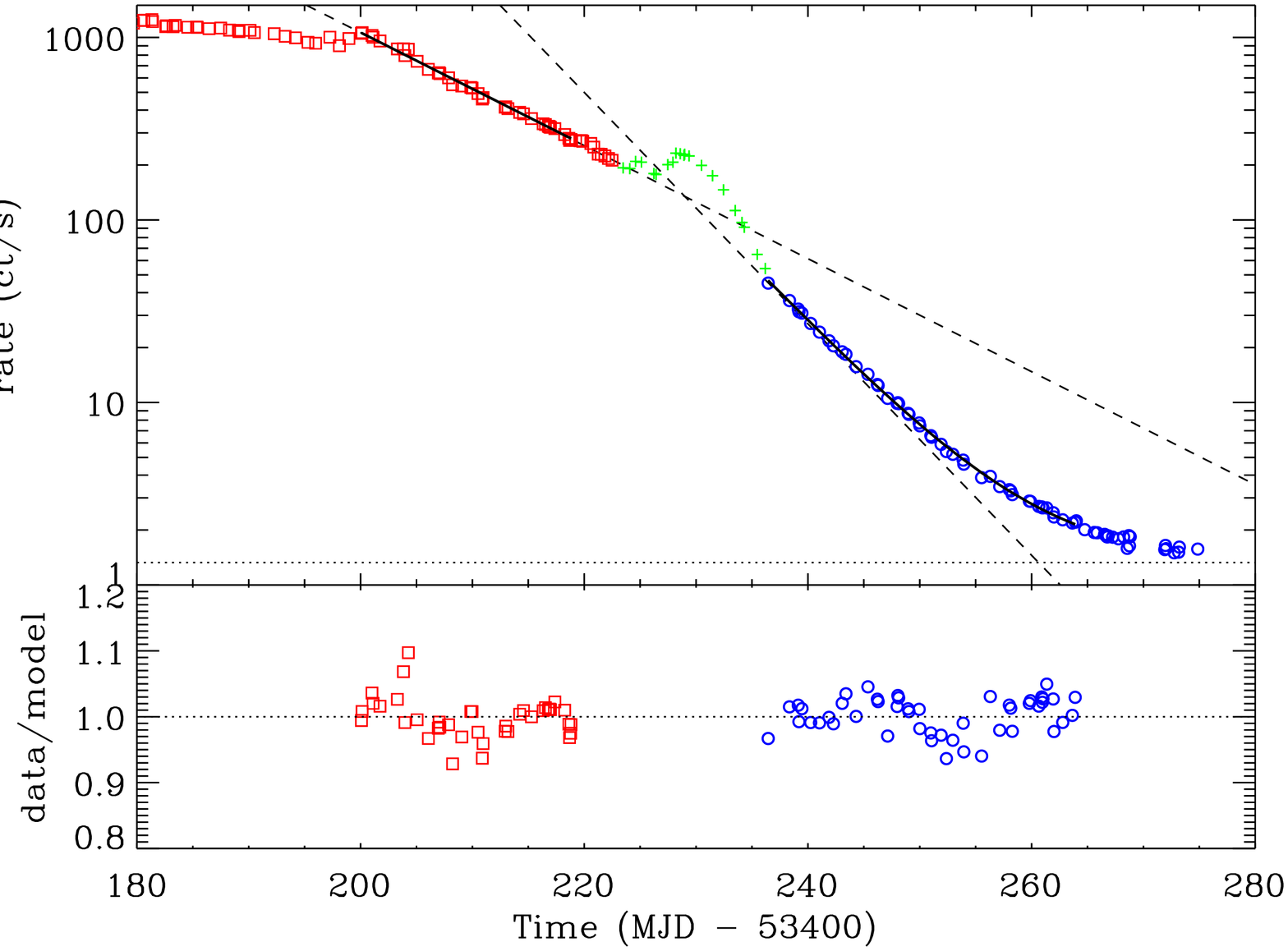}
  \caption{The decay of the 2005 outburst of \gro, fitted with separate exponential decay models in the soft state (square red points) and hard state (circular blue points). An additional constant is included in the hard-state model to account for the saturation at low fluxes (caused by background contamination). The dashed lines show the exponential decays extrapolated over the full decay region. The bottom panel shows the data/model residuals. }
  \label{fig:gro_plot_zoom}
\end{figure}

\begin{figure}
\centering
  \includegraphics[width=6.5 cm, angle=90]{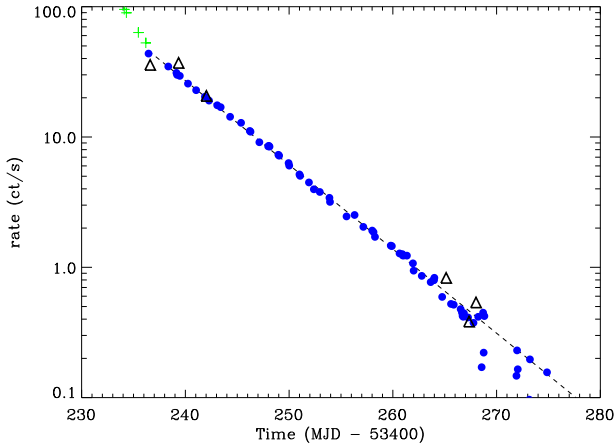}
  \caption{Close-up of the hard state decay of the 2005 outburst of \gro\ after subtracting the constant (as fitted in Fig.~\ref{fig:gro_plot_zoom}). The black triangular points show the \swift\ data, which agree with the \xte\ data after the constant has been subtracted.}
  \label{fig:gro_plot_zoom2}
\end{figure}

The clearest example of exponential decay comes from the well-observed 2005 outburst of \gro\, shown in Fig.~\ref{fig:gro_plot_zoom}. There are two distinct periods of exponential decay, ignoring the intermediate state where there is a small rise in count rate. Fitting these gave $\tau_{\rm s}$ and $\tau_{\rm h}$ values of $13.85 \pm 0.15$ and $6.72 \pm 0.08$ d, respectively. The decay rate almost exactly doubled following the transition into the hard state, giving $\beta = 1.06 \pm 0.02$. (Only the \xte\ data were fitted; the \swift\ data were used only to verify the extrapolation of the source \xte\ light curve to low fluxes.)

The other sources show more varied results with $\beta$ value ranging from $-0.1$ to $2.90$. In some cases it was not possible to clearly identify suitable decaying intervals. For some outbursts any period of exponential decay was very short, for some others it was too sparsely sampled (e.g. \xtejb). Even where there are sufficient data, in many cases the residuals for the exponential model fits show systematic structure: deviations from a single, unbroken exponential decay within a particular state. 

In order to focus on a single period of exponential decay with a constant decay rate, we used a different fitting strategy. Specifically, we calculated `local' $\tau$ values for each point by fitting only that point and the next four. We then find the modal $\tau$ and identify the longest interval in the light curve consistent with this decay rate (dominated by $\tau$ values within $1$ d of the mode). New values of $\tau_{\rm s}$, $\tau_{\rm h}$ and $\beta$ were then estimated by fitting over these smaller intervals of consistent decay rate. Furthermore, we ignore intervals containing fewer than eigth consecutive points. Table~\ref{tab:outburstdetails2} shows the results. The appendix shows the results of the fitting and provides notes on individual outbursts. This more conservative analysis yielded five $\beta$ estimates; with the exception of GX $339-4$ these lie in the range $1.0 \pm 0.3$.

\begin{table*}
  \centering
  \begin{tabular}{c c c c c c c c}
    \hline
    System name & Outburst & $\tau_{\rm s}$ (d)& Decay length & $\tau_{\rm h} $ (d)& Decay length & $\beta$ \\
    (1) & (2) & (3) & (4) & (5) & (6) & (7) \\
    \hline
    \gro\ & $2005$ & $13.85 \pm\ 0.15$ & $40$ & $6.72 \pm\ 0.08$ & $48$ & $1.06 \pm\ 0.02$ \\
    \xteja\ & $1998$ & $9.76 \pm\ 0.26$ & $15$ & $-$ & $7$ & $-$ \\
    & $1999$ & $14.80 \pm\ 1.11$ & $8$ & $6.95 \pm\ 0.08$ & $22$ & $1.13 \pm\ 0.09$ \\
    \xtejb\ & $2001$ & $16.12 \pm\ 0.27$ & $20$ & $-$ & $-$ & $-$ \\
    \gx\ & $2002$ & $12.04 \pm\ 0.45$ & $8$ & $8.46 \pm\ 1.24$ & $9$ & $0.42 \pm\ 0.06$ \\
    & $2004$ & $-$ & $4$ & $8.78 \pm\ 0.70$ & $21$ & $-$ \\
    & $2007$ & $-$ & $7$ & $8.87 \pm\ 1.32$ & $16$ & $-$ \\
    & $2010$ & $-$ & $-$ & $5.88 \pm\ 0.68$ & $8$ & $-$\\
    \fouru\ & $2002$ & $6.10 \pm\ 0.15$ & $12$ & $2.70 \pm\ 0.08$ & $12$ & $1.26 \pm\ 0.05$\\
    \h\ & $2003$ & $12.96 \pm\ 0.51$ & $8$ & $7.41 \pm\ 0.91$ & $14$ & $0.75 \pm\ 0.10$ \\
    & $2004$ & $-$ & $6$ & $5.19 \pm\ 0.27$ & $13$ & $-$\\
    & $2005$ & $-$ & $6$ & $-$ & $6$ & $-$ \\
    & $2009$ & $-$ & $7$ & $-$ & $6$ & $-$ \\
    & $2010$ & $-$ & $6$ & $10.69 \pm\ 2.40$ & $10$ & $-$ \\
    & $2011$ & $-$ & $-$ & $-$ & $7$ & $-$ \\
    \hline
  \end{tabular}
  \caption{The same parameters as in Table~\ref{tab:outburstdetails} after fitting only restricted sections of the light curves (see the text for details).}
  \label{tab:outburstdetails2}
\end{table*}

%Specifically, we began by fitting just the five last data points of the soft state, and the first five data points of the hard state. From each of these fits we computed the $\chi^2$ and corresponding $p$-value. We then included progressively more data points until there was a large (factor $>5$) drop in the $p$-value, indicating a significant deviation from an initially exponential decay. Columns $4$--$8$ of Table~\ref{tab:outburstdetails2} show the results obtained using intervals defined in this way, i.e. the longest intervals consistent with a single exponential decay immediately before or after the soft-hard state transition. These show much closer agreement to $\beta \approx 1$, and typically $\tau_s \approx 2$ weeks and $\tau_h \approx 1$ week.

% \begin{figure*}
% \centering
%  \includegraphics[width=12cm]{combinedlcbkg.ps}
%  \caption{Decaying sections of the light curve of each of the outbursts, arranged to fit on the same scale. The start times of each decay have been set to zero.}
%  \label{fig:plot_lc}
% \end{figure*}

%______________________________________________________________

\section{Discussion}
\label{sect:disco}

We have examined the decaying stages in the X-ray light curves of 15 outbursts of six BH X-ray binaries. We found that in all cases the last few weeks of each outburst can be described in terms of nearly exponential decays. For the seven outbursts with well-observed exponential decays in the soft state, the mean e-folding time is $12 \pm 1$ d, while for the ten outbursts with well observed exponential decays in the hard state, the mean e-folding time is $7 \pm 1$ d (see Table 2). Several outbursts (e.g. of \gx) show more complicated behaviour, sometimes with little or no obvious decay during the soft state. As such, it is difficult to say whether the models discussed above will apply to all BH LMXB systems/outbursts or just a subset. \citet{homanAtel2005} previously noted the factor 2 change in decay time-scale in their early analysis of the 2005 outburst of \gro. Here, we have confirmed their results for that one outburst and showed that similar behaviour can be seen in the other four outbursts with well-constrained exponential decays in both states.

We characterized the change in the decays during soft and hard states using a parameter $\beta  = \tau_{\rm s}/\tau_{\rm h} - 1$. Assuming constant $\eta$ in the soft state, such that $\tau_{\rm s} = \tau_{\rm m}$ (the decay of $\dot{M}$), this parameter can be interpreted as the index of the efficiency scaling: $\eta \propto \dot{M}^{\beta}$. Of the five systems with the most reliable $\beta$ estimates, four show $\beta \approx 1$, while \gx\ shows $\beta \approx 0.4$.

%The exceptions are the 2007 outburst of \gx\ and the outbursts of \h; these display complicated soft state decays and/or are rather sparsely sampled.

% {\bf [Graham: based just on the $\tau$ we measure can we say something about the disc size and/or viscosity? It's a lot shorter than King \& Ritter's 40d timescale!]}

We estimated the decay rates using broad-band light curves ($3$-$20$ keV) from the \xte\ PCA. Such a dramatic change in the decay rate is not an artifact of spectral evolution: within each of the soft and hard states, the spectral evolution is relatively modest (see Fig.\ref{fig:gro_plot}). We conclude that there is a genuine change in the speed at which the X-ray luminosity drops. If $L_{\rm X} = \eta \dot{M} c^2$, then either the decay rate of $\dot{M}$ changes around the state transition or $\eta$ decreases with time (and luminosity) after the state transition.

The change in the decay could signal the switch from a disc-dominated, radiatively efficient soft state to a radiatively inefficient mode of accretion (such as an ADAF or JDAF type flow) in the hard state. As discussed above, such a model -- in which the soft state has $\eta$ constant with $\dot{M}$, but the hard state is governed by an accretion flow that becomes radiatively inefficient at low $\dot{M}$ according to $\eta \propto \dot{M}$ -- matches the results. This interpretation is consistent with other studies comparing radio and X-ray luminosities over a sample of binaries \citep{fender03, merloni03}. By contrast, \cite{knevitt14} required a steeper efficiency scaling (with $\beta > 3$) to explain the lack of known short period BH LMXBs relative to NS LMXBs. Reconciling these results -- the factor $\approx 2$ change in the decay slopes implying $\eta \propto \dot{M}$ scaling, and the lack of short period BH LMXBs -- would require either that there is another explanation for the absence of short-period BH systems, or that the efficiency scaling becomes much steeper at lower luminosities, below those accessible with the \xte\ data.

It is the average radiative efficiency (averaged over any contributing spectral components) that is required to change. There could be two emission components with radiative efficiencies that are very different, but each is constant with time. The observed faster decay in the hard state could be produced if the source emission evolves from being dominated by the more radiatively efficient component to being dominated by the least radiatively efficient component, so that the weighted average of the radiative efficiency drop with $L_{\rm X}$. But such a model requires fine-tuning to explain the observed results. Consider two emission processes with spectra $S_1(E)$ and $S_2(E)$, and radiative efficiencies $\eta_1$ and $\eta_2$ (with $\eta_2 < \eta_1$). The mass accretion rate $\dot{M}$ powering the source is divided such that a fraction $a$ drives the first component and $(1-a)$ drives the second. The total luminosity is $L \propto \dot{M}(\eta_1 a S_1 + \eta_2 (1-a) S_2)$. The term in brackets is the emission-averaged radiative efficiency. In order to reproduce the doubling in decay rate, $a$ would have to be a function of $\dot{M}$ in the hard state in such a way to make the emission-averaged radiative efficiency scale linearly with $\dot{M}$, which seems contrived. Further, the relatively modest evolution of the spectrum during the hard-state decay does not support such a model unless the two components have very similar spectra over the observed X-ray band.

The above interpretation assumes that the mass accretion rate decays as $\dot{M} \propto \exp(-t/\tau_{\rm m})$ through most of the outburst decay, and in particular, $\tau_{\rm m}$ does not change around the state transition. An alternative explanation is that $\eta$ is constant and there is a change in $\tau_{\rm m}$ coincident with the state transition. The coincidence in time would require that the state transition and the change in the decay rate of $\dot{M}$ are causally linked (either one causes the other or they are both triggered by some other event) by some mechanism not included in the current models of disc outbursts \citep{vanparadijs96, king98, dubus01, lasota01}

We conclude that the most plausible explanation is in terms of a change in the radiative efficiency, and that the soft- to hard-state transition is the result of the inner accretion flow transforming from a standard, efficient accretion disc to a radiatively inefficient flow as suggested by models such as \citet{narayan95} and \citet{esin97}.

%______________________________________________________________

\section*{Acknowledgements}

AJE is supported by an STFC studentship. SV and GAW acknowledge support from STFC consolidated grant ST/K001000/1. This research has made use of NASA's Astrophysics Data System and of data, software and web tools obtained from NASA's HEASARC, a service of Goddard Space Flight Center and the Smithsonian Astrophysical Observatory. We thank an anonymous referee for a careful reading of the paper and a constructive report.

%______________________________________________________________

\bibliographystyle{mn2e}
\bibliography{references}

\bsp

\vfill
\pagebreak

\appendix
\section{Additional Plots}

\begin{figure}
\centering
  \includegraphics[width=6.0cm, angle=90]{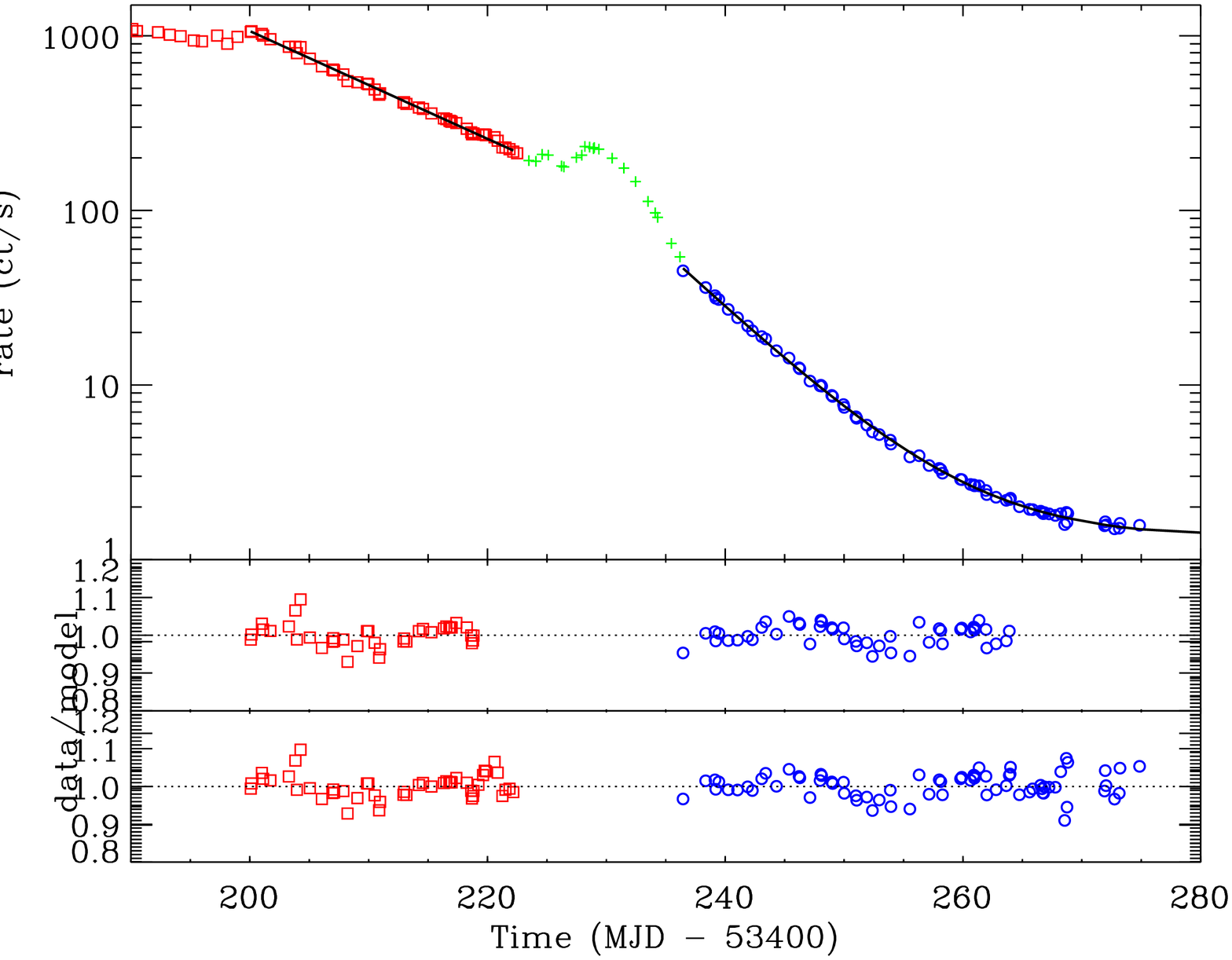}
  \caption{The decaying stages of the 2005 outburst of \gro. The top panel shows the $3$-$20$ keV light curve with exponential decay models (black) fitted to both soft-state (red squares) and hard-state (blue circles) data. The bottom panel shows the residuals from the fit described in Table 1, the middle panel shows the results from the fit described in Table 2, using only intervals of nearly uniform decay. (A constant component was added to the model as described in the text.)}
  \label{fig:gro_plot_residuals}
\end{figure}

\begin{figure}
\centering
  \includegraphics[width=6.0cm, angle=90]{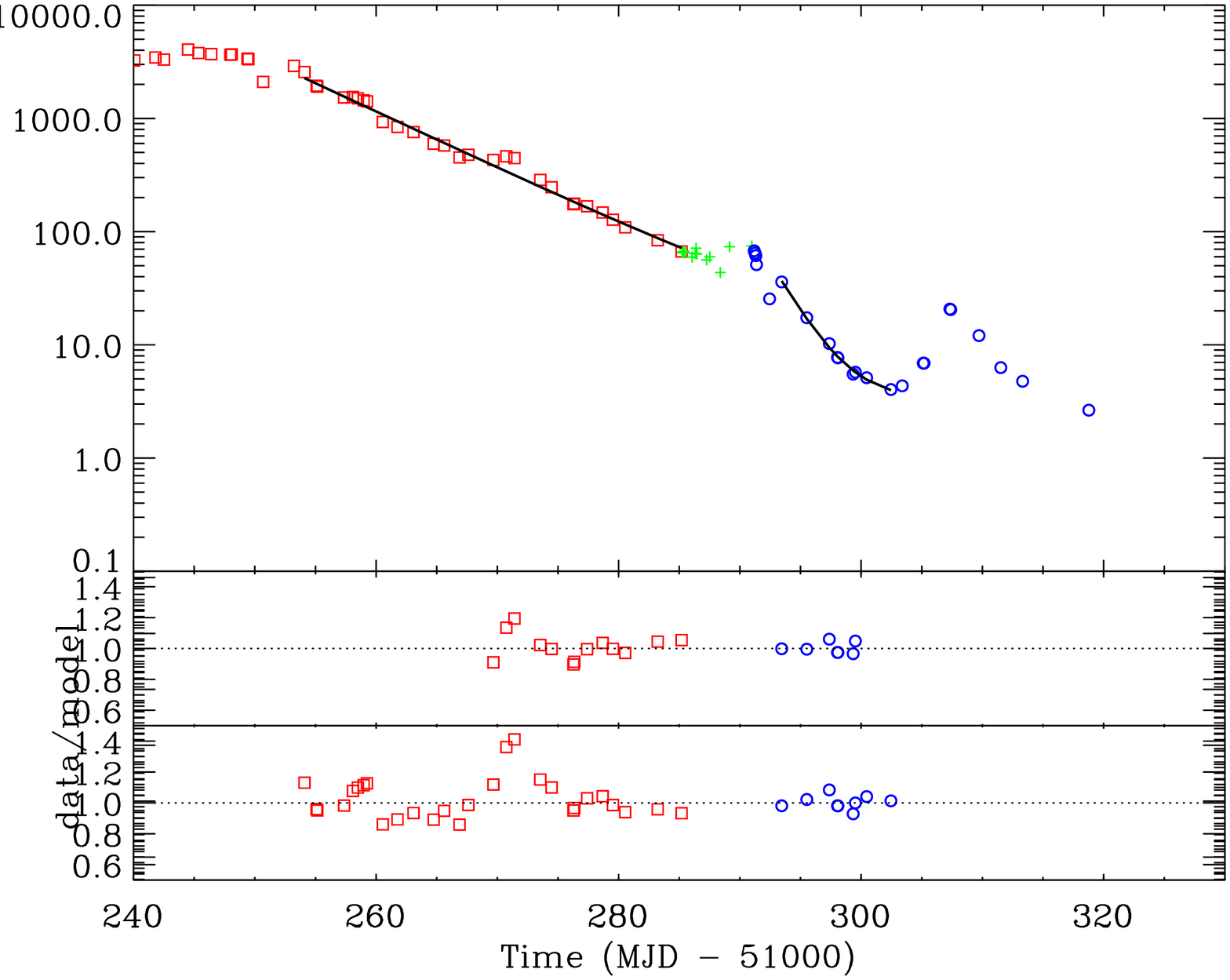}
  \caption{The decaying stages of the 1998 outburst of \xteja. The soft-state has a more complicated decay that is not a single exponential decay due to the flare around MJD $51270$. The hard state itself is also affected by a flare, this time dominating the whole decay, leaving only a small section that could be fit.}
  \label{fig:xtej1_plot_residuals}
\end{figure}

\begin{figure}
\centering
  \includegraphics[width=6.0cm, angle=90]{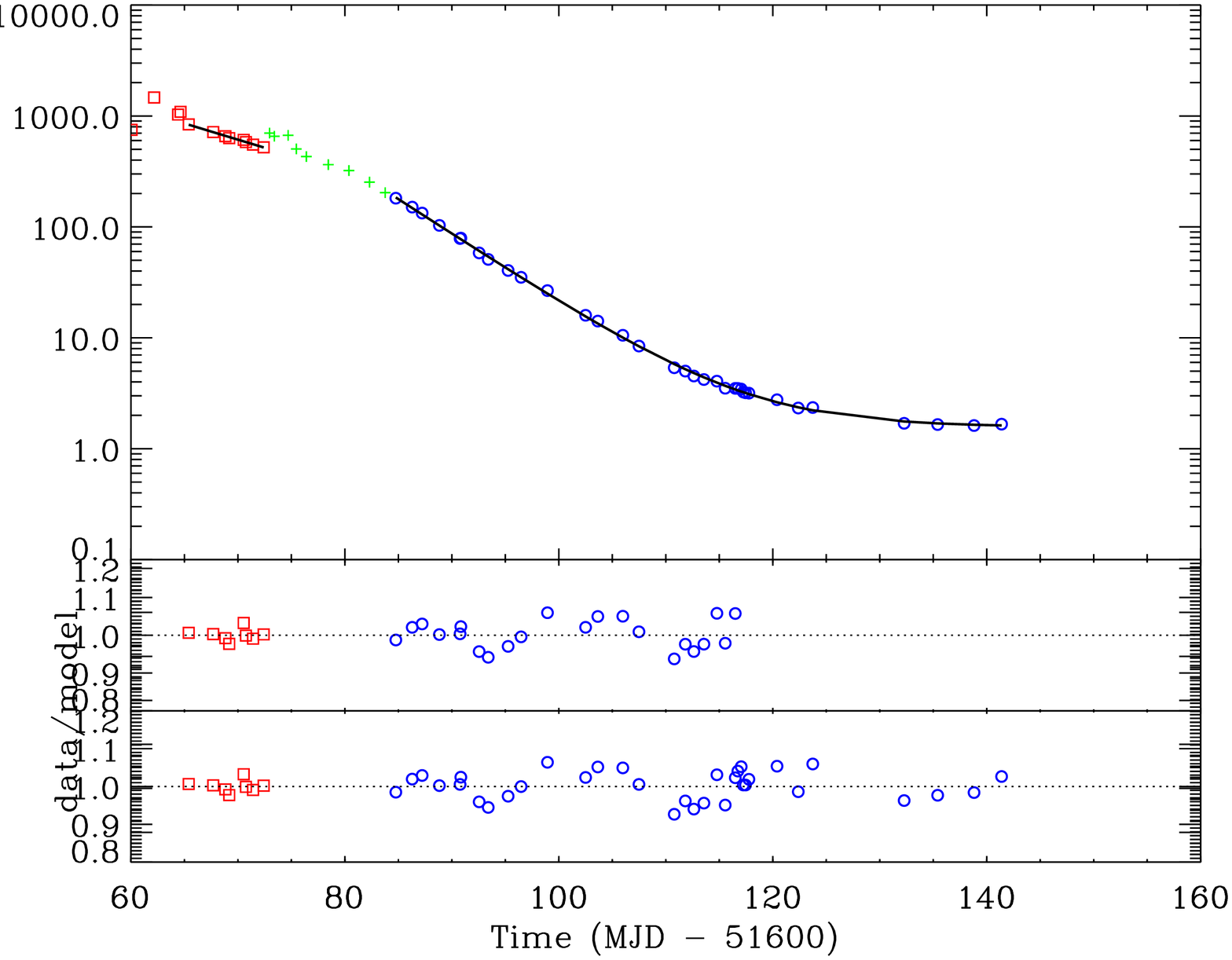}
  \caption{The decaying stages of the 1999 outburst of \xteja.}
  \label{fig:xtej2_plot_residuals}
\end{figure}

\begin{figure}
\centering
  \includegraphics[width=6.0cm, angle=90]{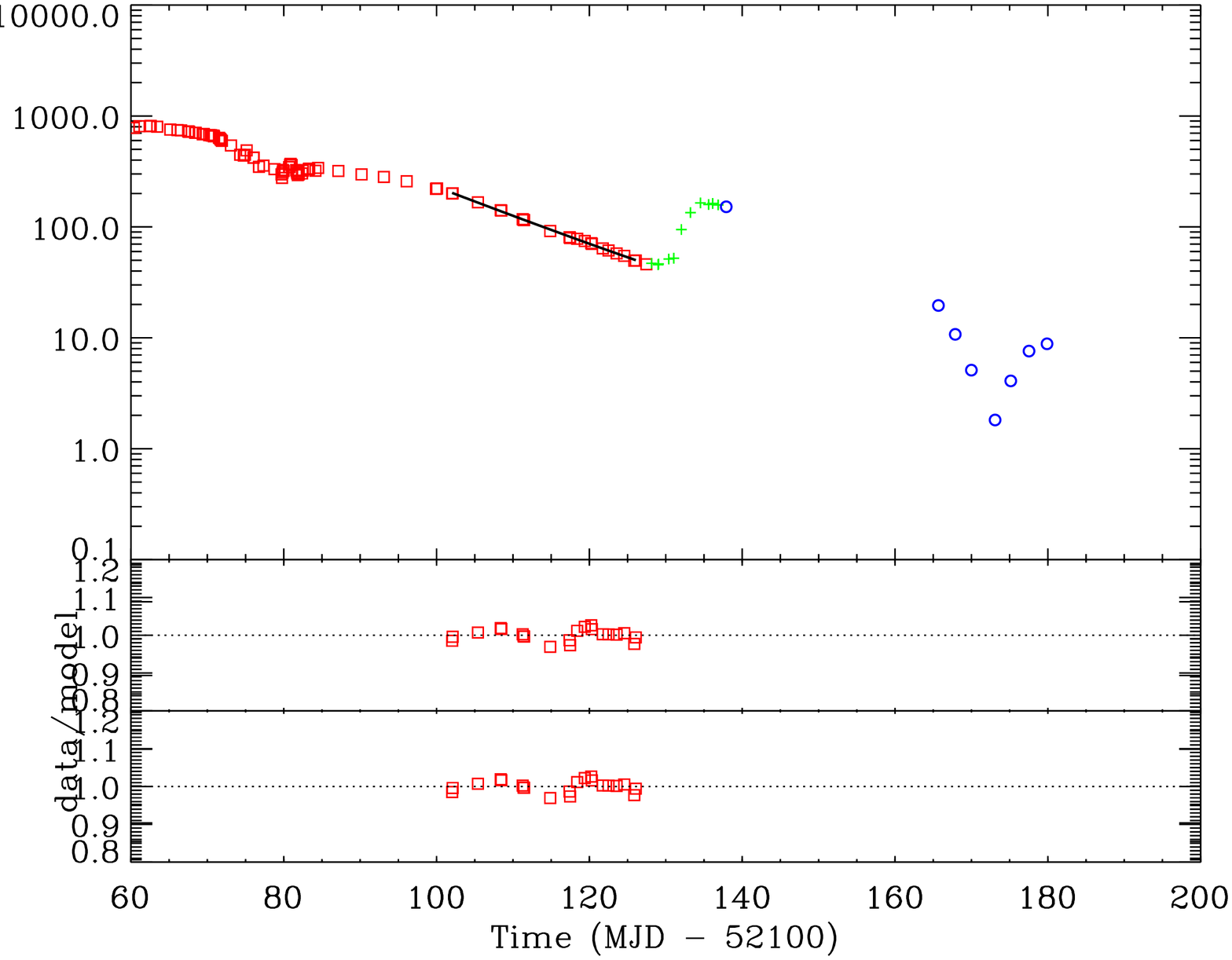}
  \caption{The decaying stages of the 1999 outburst of \xtejb.}
  \label{fig:xtej3_plot_residuals}
\end{figure}

\begin{figure}
\centering
  \includegraphics[width=6.0cm, angle=90]{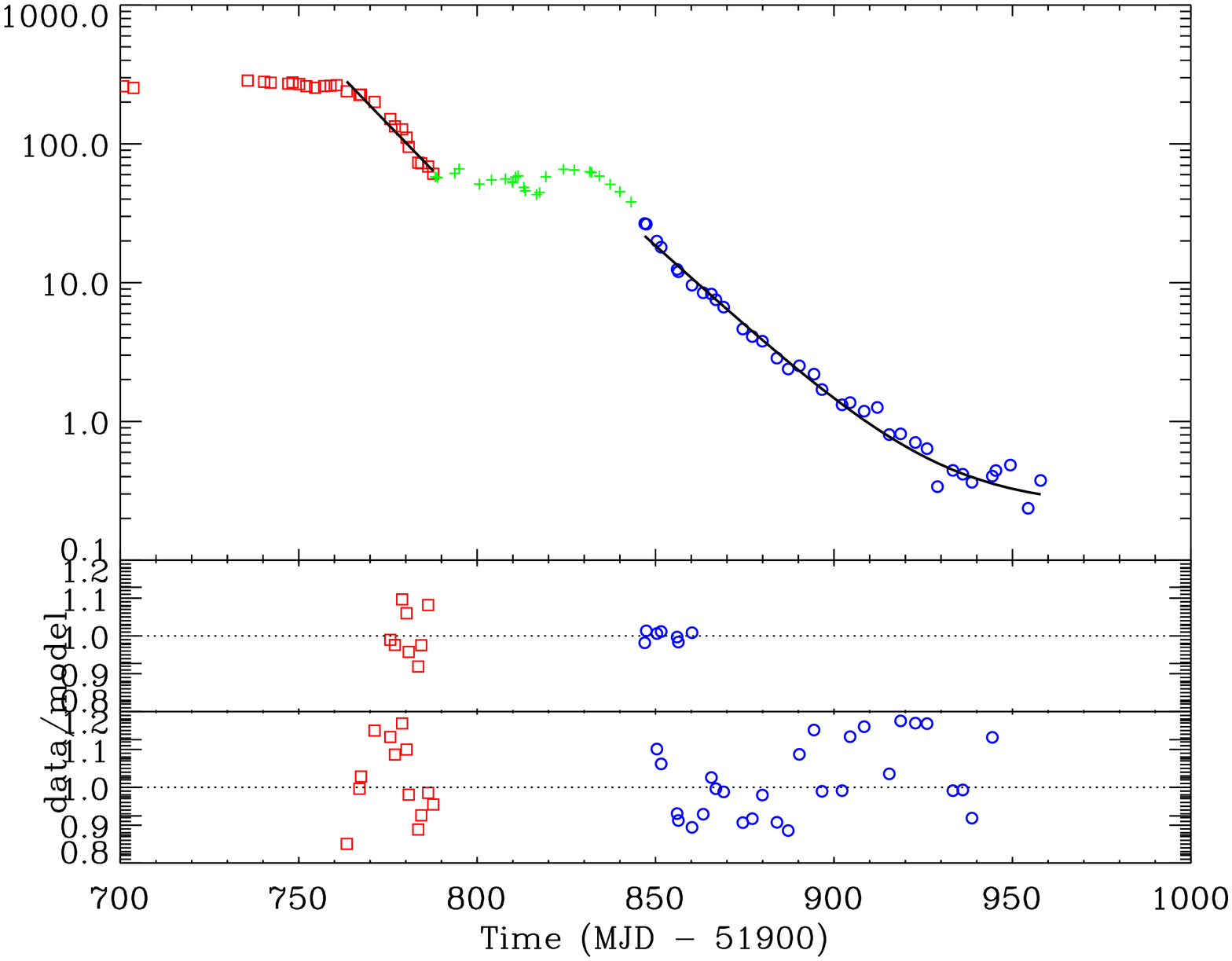}
  \caption{The decaying stages of the 2002 outburst of \gx. We have $\beta$ for this outburst, however there are certain features of the \gx\ outbursts that should be noted. Compared to the overall length of the soft state, the soft decaying interval is very short, with the source staying at near constant luminosity for many weeks.}
  \label{fig:gx1_plot_residuals}
\end{figure}

\begin{figure}
\centering
  \includegraphics[width=6.0cm, angle=90]{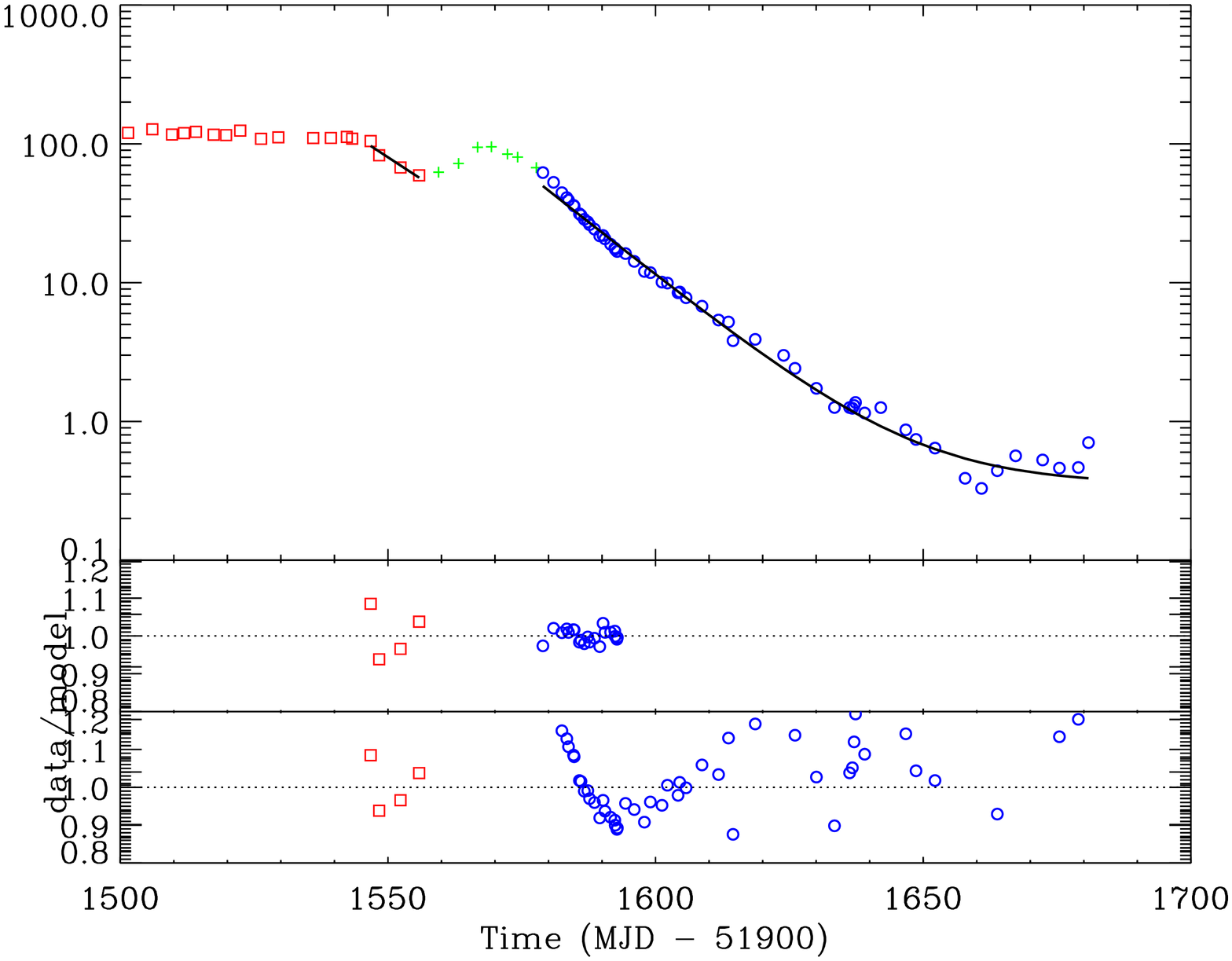}
  \caption{The decaying stages of the 2004 outburst of \gx. The soft-state decay only lasts for around a week, making it too short to reliably measure its properties.}
  \label{fig:gx2_plot_residuals}
\end{figure}

\begin{figure}
\centering
  \includegraphics[width=6.0cm, angle=90]{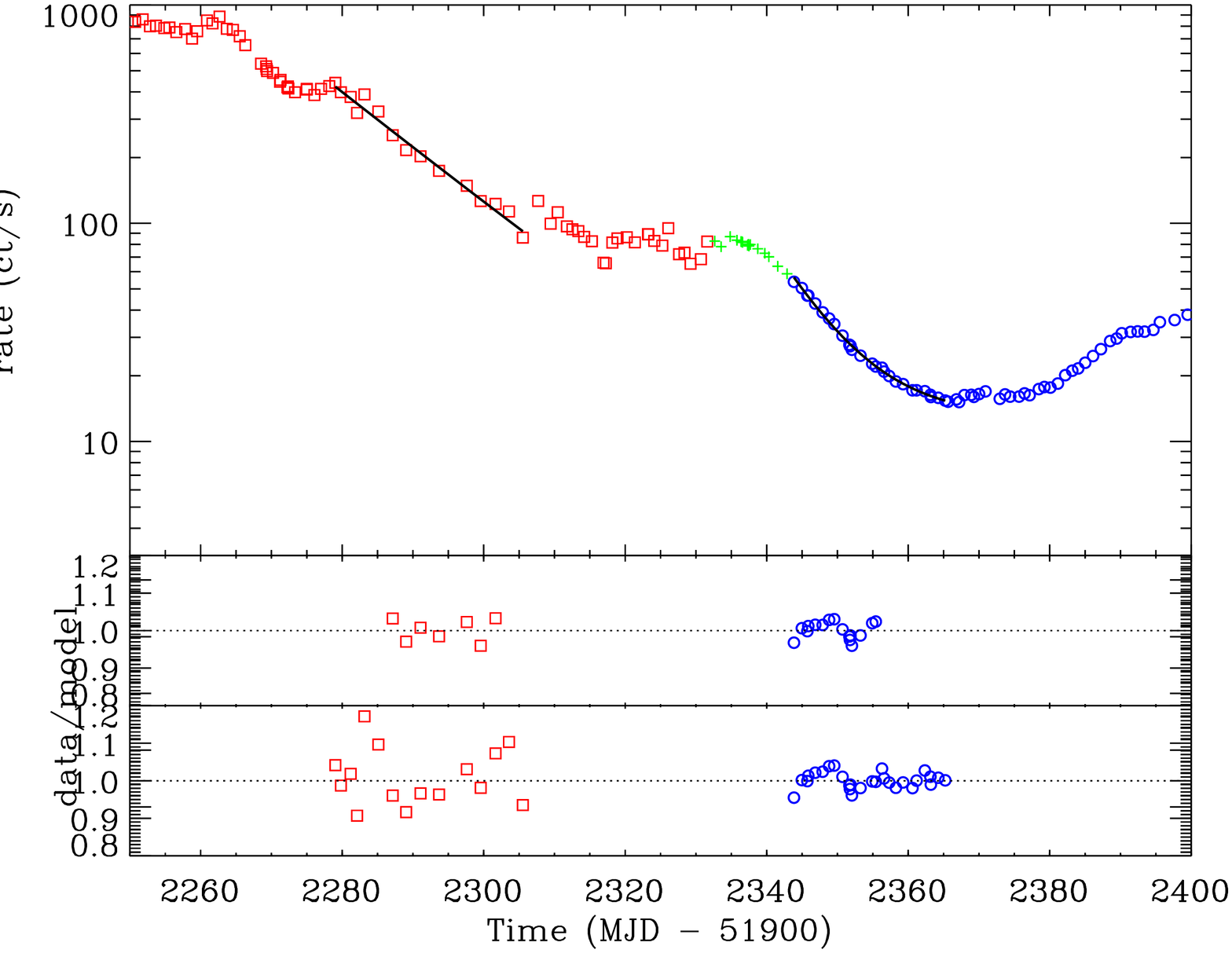}
  \caption{The decaying stages of the 2007 outburst of \gx.}
  \label{fig:gx3_plot_residuals}
\end{figure}

\begin{figure}
\centering
  \includegraphics[width=6.0cm, angle=90]{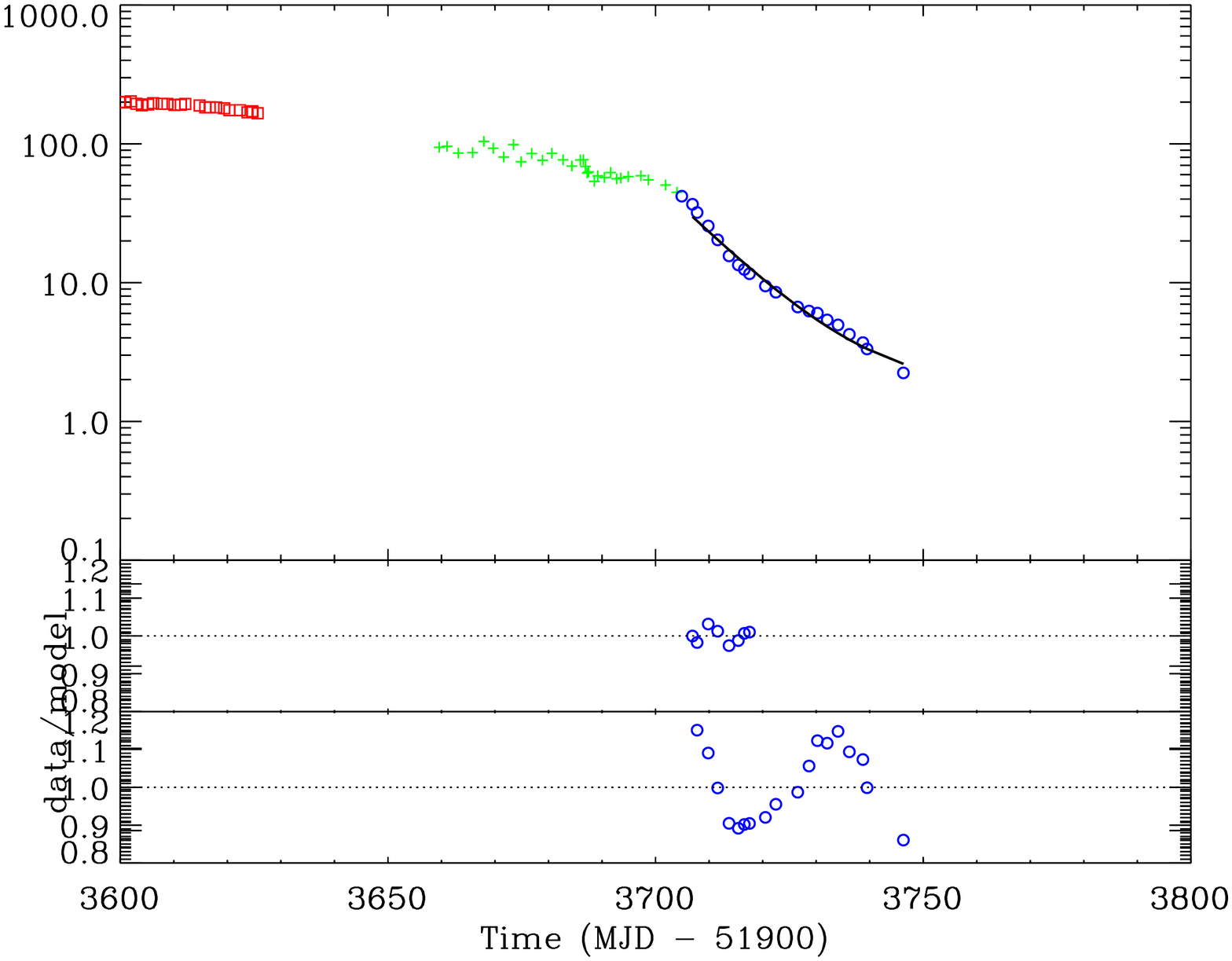}
  \caption{The decaying stages of the 2010 outburst of \gx.}
  \label{fig:gx4_plot_residuals}
\end{figure}

\begin{figure}
\centering
  \includegraphics[width=6.0cm, angle=90]{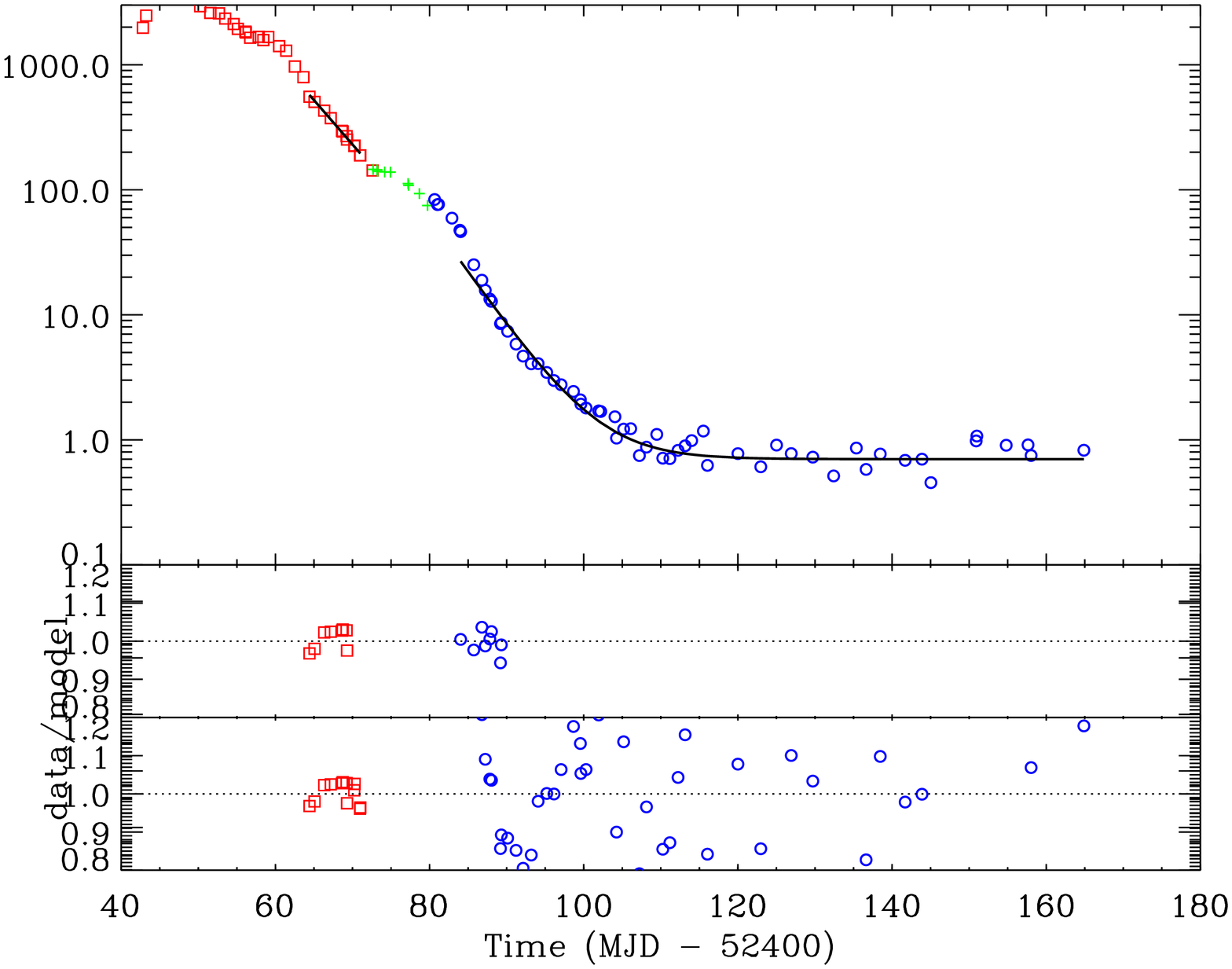}
  \caption{The decaying stages of the 2002 outburst of \fouru.}
  \label{fig:4u_plot_residuals}
\end{figure}

\begin{figure}
\centering
  \includegraphics[width=6.0cm, angle=90]{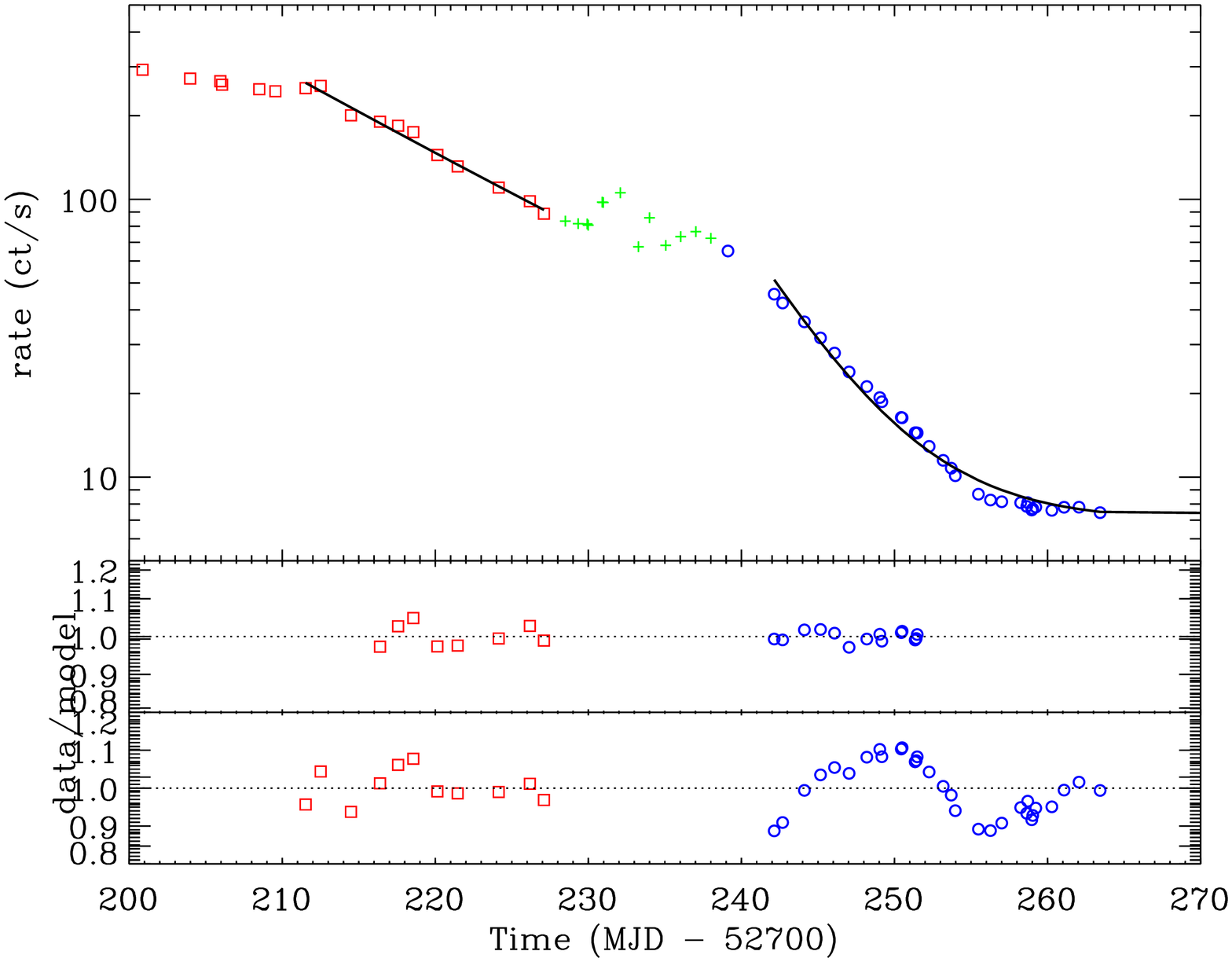}
  \caption{The decaying stages of the 2003 outburst of \h.}
  \label{fig:h1_plot_residuals}
\end{figure}

\begin{figure}
\centering
  \includegraphics[width=6.0cm, angle=90]{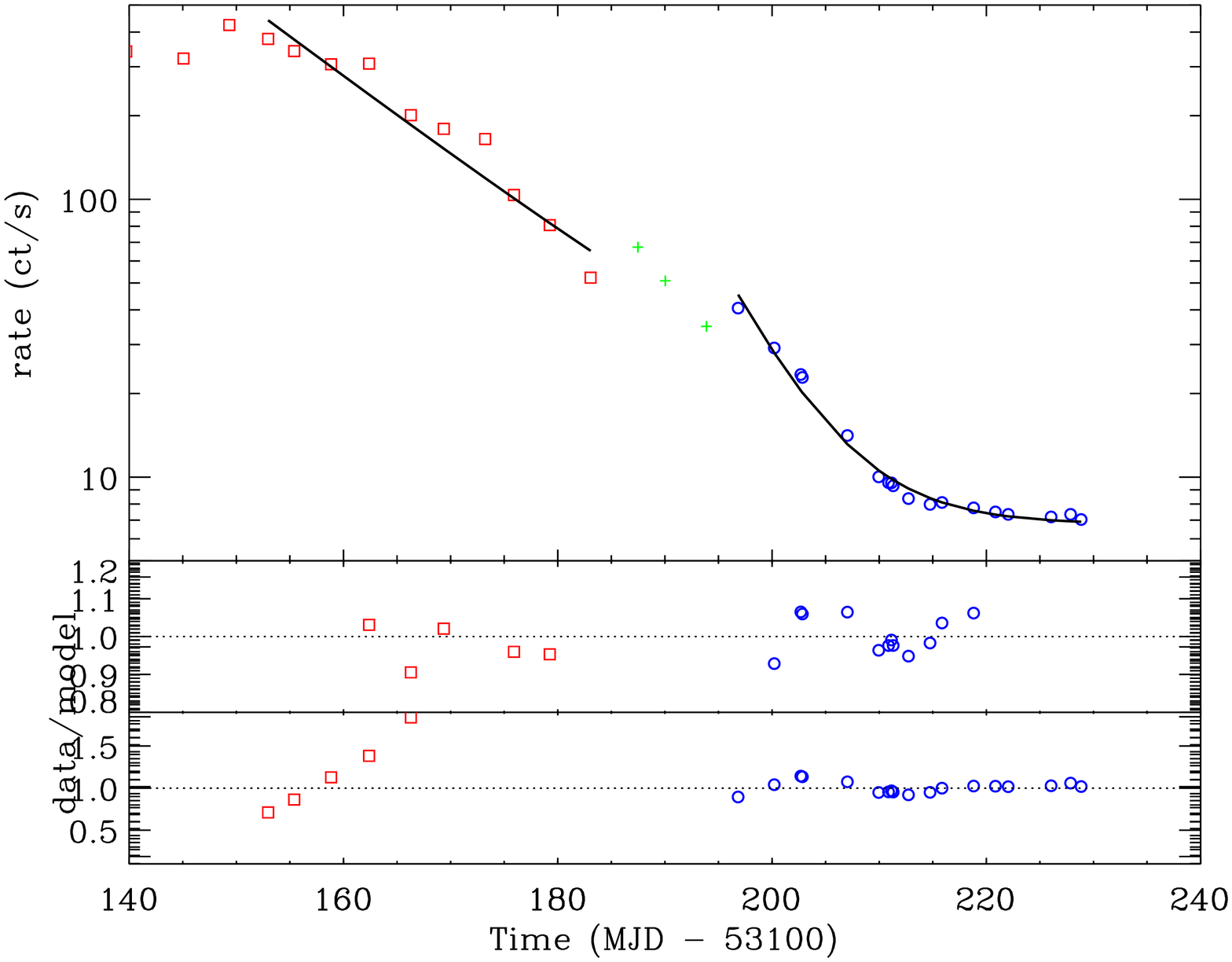}
  \caption{The decaying stages of the 2004 outburst of \h.}
  \label{fig:h2_plot_residuals}
\end{figure}

\begin{figure}
\centering
  \includegraphics[width=6.0cm, angle=90]{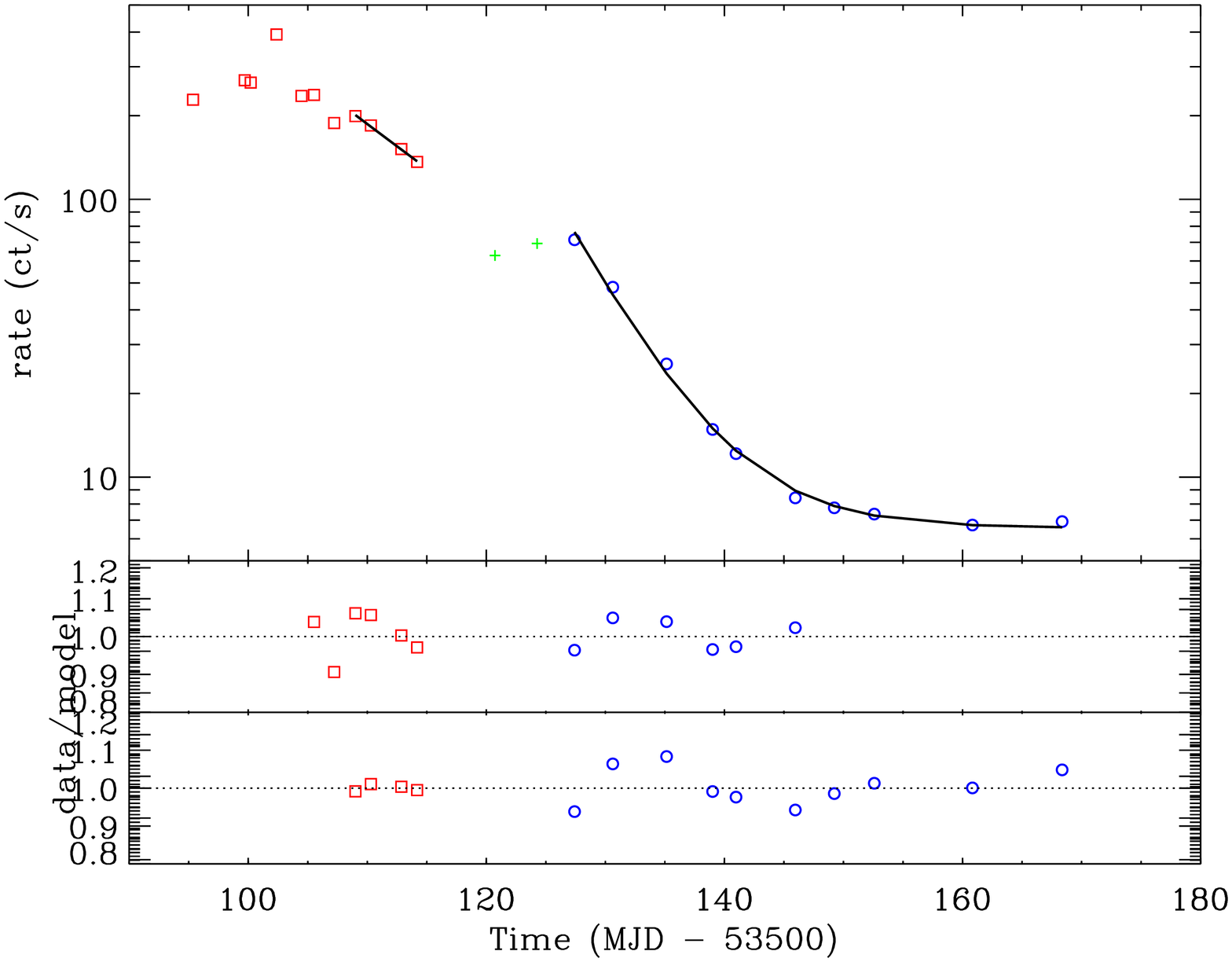}
  \caption{The decaying stages of the 2005 outburst of \h.}
  \label{fig:h3_plot_residuals}
\end{figure}

\begin{figure}
\centering
  \includegraphics[width=6.0cm, angle=90]{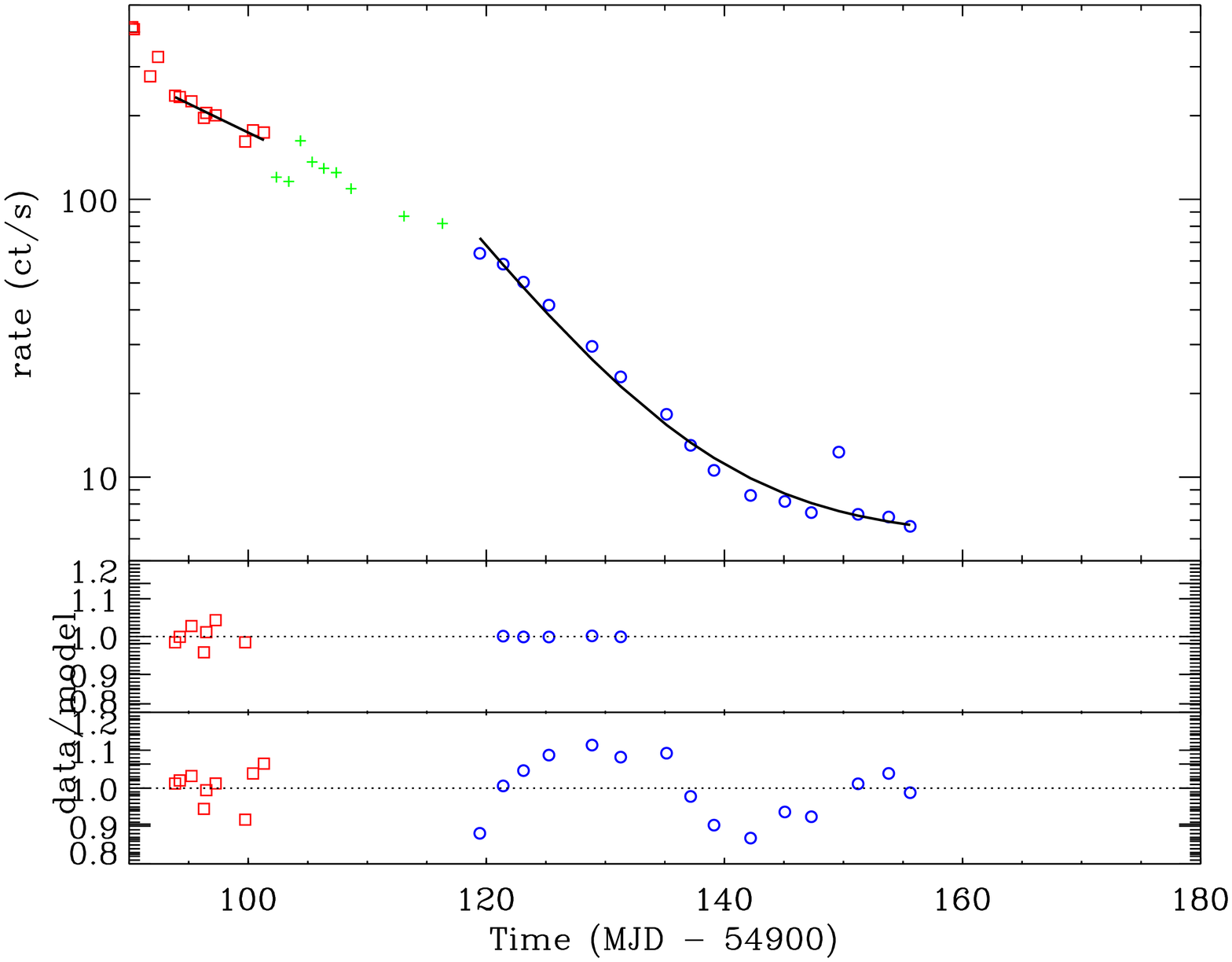}
  \caption{The decaying stages of the 2009 outburst of \h.}
  \label{fig:h4_plot_residuals}
\end{figure}

\begin{figure}
\centering
  \includegraphics[width=6.0cm, angle=90]{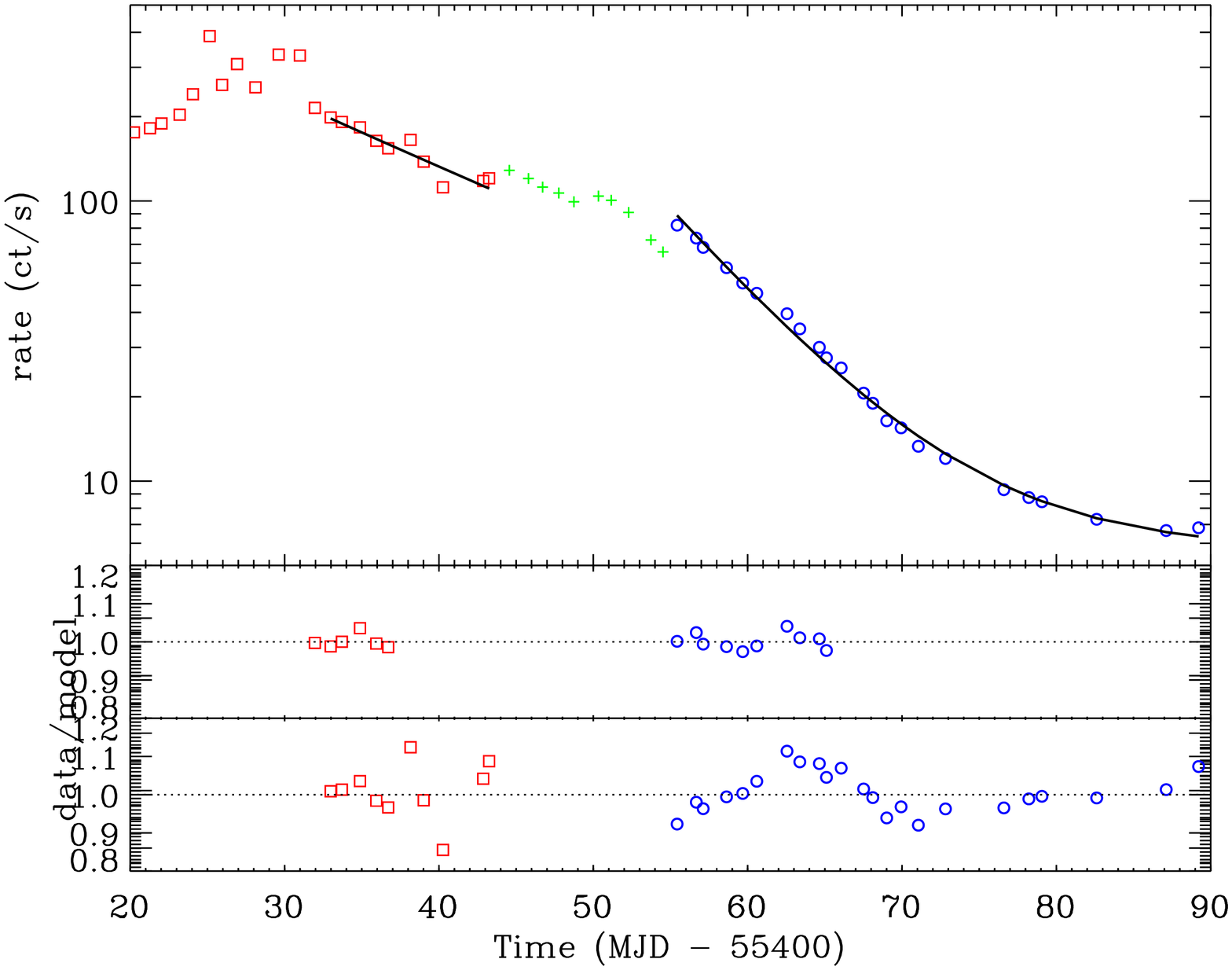}
  \caption{The decaying stages of the 2010 outburst of \h.}
  \label{fig:h5_plot_residuals}
\end{figure}

\begin{figure}
\centering
  \includegraphics[width=6.0cm, angle=90]{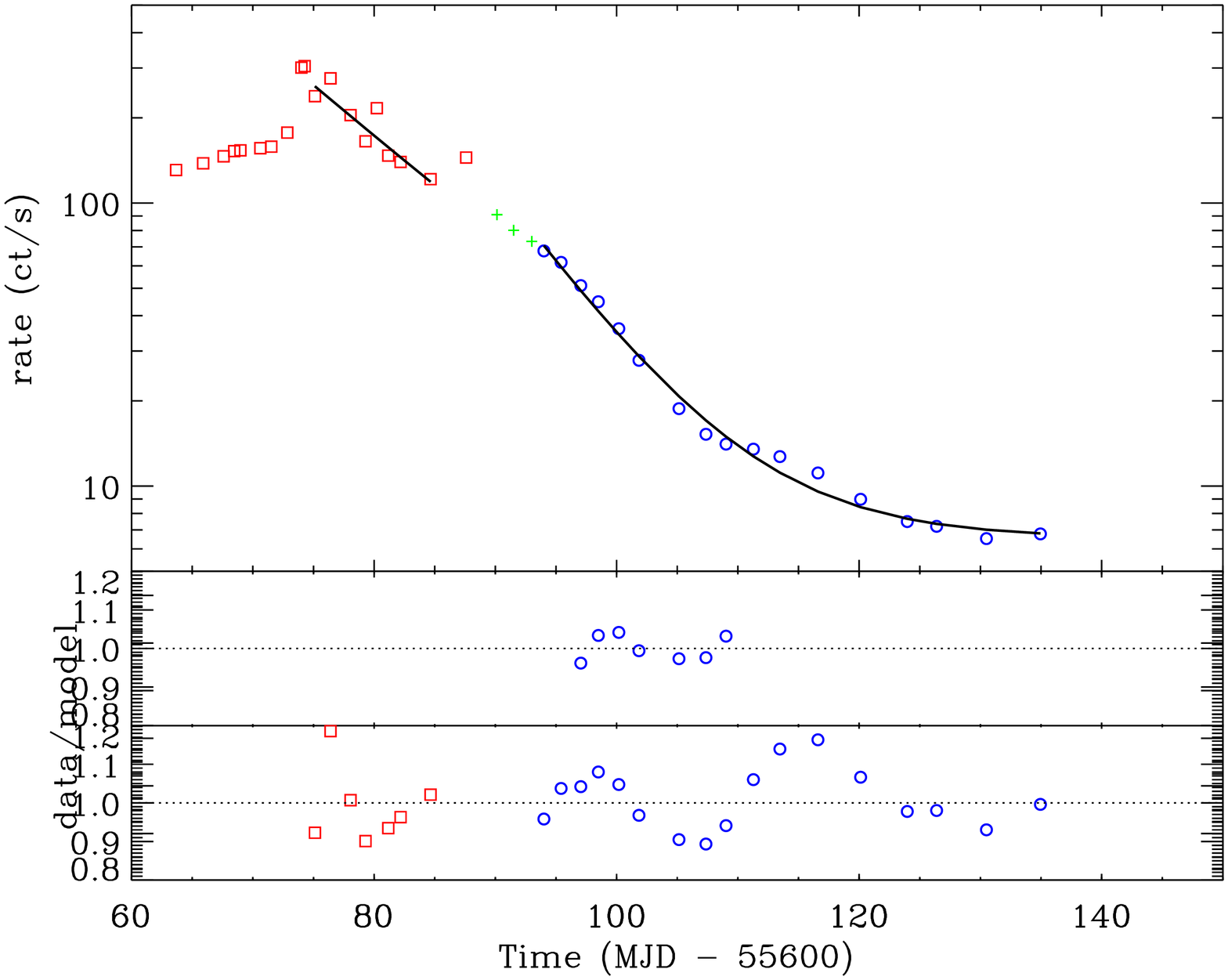}
  \caption{The decaying stages of the 2011 outburst of \h.}
  \label{fig:h6_plot_residuals}
\end{figure}

\label{lastpage}

\end{document}